\documentclass[a4paper,preprintnumbers,article,amsmath,amssymb,floatfix,10pt,prd,superscriptaddress,nofootinbib,twocolumn]{revtex4}

\usepackage{doi}
\usepackage{hyperref}
\hypersetup{
  colorlinks=true,        
  linkcolor=blue,         
  citecolor=red,         
}

\usepackage{graphicx}
\usepackage{dcolumn}
\usepackage{bm}
\usepackage{color}
\usepackage{enumitem}
\usepackage{amsmath}
\usepackage{amssymb}
\usepackage{orcidlink}

\newcommand{\beq}{\begin{equation}}
\newcommand{\eeq}{\end{equation}}
\newcommand{\bea}{\begin{eqnarray}}
\newcommand{\eea}{\end{eqnarray}}

\usepackage{bbm}
\usepackage{amsfonts}
\usepackage{mathrsfs}
\usepackage{latexsym}
\usepackage{epsfig}
\usepackage{epstopdf}
\usepackage{epstopdf}
\usepackage{graphicx}
\usepackage{amssymb}
\usepackage{amsmath}
\usepackage{dcolumn}
\usepackage{bm}
\usepackage{color}
\usepackage{comment}
\usepackage{xcolor}
\usepackage{float, geometry,lipsum}
\begin{document}

\title{Gravitational dynamics and backward ray-traced shadows of bumblebee AdS black holes} 
 
\author{Muhammad Yasir}
\email{yasirciitsahiwal@gmail.com; muhammadyasir@nbu.edu.cn}
\affiliation{Institute of Fundamental Physics and Quantum Technology, Department of Physics, School of Physical Science and Technology, Ningbo University, Ningbo, 315211, Zhejiang, People’s Republic of China.}
\author{Senyue Lou}
\email{lousenyue@nbu.edu.cn (corresponding author)}
\affiliation{Institute of Fundamental Physics and Quantum Technology, Department of Physics, School of Physical Science and Technology, Ningbo University, Ningbo, 315211, Zhejiang, People’s Republic of China.}
\author{Zelin Zhang}
\email{zhangzelin1@nbu.edu.cn}
\affiliation{Institute of Fundamental Physics and Quantum Technology, Department of Physics, School of Physical Science and Technology, Ningbo University, Ningbo, 315211, Zhejiang, People’s Republic of China.}
\author{Abdul Jawad}
\email{abduljawad@cuilahore.edu.pk}
\affiliation{Department of Mathematics, COMSATS University Islamabad, Lahore-Campus, Lahore, 54000, Pakistan.}
\author{Faren Wang}
\email{fr.wang@outlook.com}
\affiliation{Institute of Fundamental Physics and Quantum Technology, Department of Physics, School of Physical Science and Technology, Ningbo University, Ningbo, 315211, Zhejiang, People’s Republic of China.}

\begin{abstract}
We investigate the total observed frequency shift of photons emitted by massive geodesic particles in circular orbits near the event horizon of bumblebee AdS black holes (BHs). The cosmological effects of the bumblebee field are investigated through dynamical equations governing the evolution and characteristic features of the Universe. In addition, the bumblebee field generates cosmological anisotropies by acting as a source of anisotropy and selecting a preferred spatial direction. 
We formulate the characteristic equations for the innermost stable circular orbit (ISCO) using the effective potential approach. Further analysis is devoted to establishing a functional relationship between the BH mass and the observed redshift and blueshift of photons emitted from the accretion disk. Finally, we study the shadow cast by the bumblebee AdS BH using the backward ray-tracing process implemented in the Arcmancer code, where photon geodesics are initialized at the observer’s location and integrated backward in time to generate BH images, typically illuminated by electromagnetic radiation from nearby accretion flows.
\end{abstract}

\maketitle

\section{Introduction}

Black holes are fundamental solutions of Einstein’s field equations and are considered robust predictions of General Relativity (GR). Their existence has been validated through several observational campaigns, despite historical debates. The observation of the bending of light by gravitational lensing is a crucial tool in cosmology  to examine the planets, dark energy, and dark matter.
Light bending possesses significant theoretical importance, particularly in probing the null structure of spacetime. Gravitational lensing can also be used to evaluate the mass of astrophysical objects located between the light source and  an observer \cite{l1,l2}.
The Event Horizon Telescope (EHT) provided the first direct observational confirmation of the BH shadow at the center of the galaxy \(M87\) \cite{l3}. Furthermore, recent advances in EHT observations have enabled polarization imaging of \(M87^{*}\), revealing the magnetic field structure near the BH and providing important insight into the physical mechanisms responsible for jet formation \cite{l5}.

The existence of the ISCO is widely regarded as one of the fundamental strong-gravity effects in BH accretion disk theory. Nevertheless, the physical role of the ISCO remains a subject of ongoing debate. According to one viewpoint \cite{IS1}, for sufficiently low accretion rates, the ISCO provides the effective “inner edge” of the accretion disk. It separates the outer region, where matter follows nearly Keplerian and quasi-circular orbits from the plunging region, in which matter falls almost freely into the BH.
In particular, this method is used to estimate BH spin through spectral fitting, which relies on this assumption \cite{IS2}. In contrast, an alternative viewpoint, most recently summarized by Balbus \cite{IS3}, argues that the ISCO does not play a dynamically significant role in BH accretion, even at low accretion rates. In this picture, magnetohydrodynamic turbulence driven by the magnetorotational instability (MRI) operates on both sides of the ISCO, rendering the accretion flow unstable and turbulent throughout the inner disk region.

Recent discoveries in the detection of BH nature have led astrophysicists to design innovative strategies to determine BH parameters through various types of measurements, thereby obtaining deeper insights into fundamental physics \cite{l11,l12,l13,l13B1, l13B2,l13B3}.
This observational data permitted the aforementioned teams to estimate the putative BH mass of Sgr~A* and the distance from Earth to its center, yielding the values \( M\sim 4.3 \times 10^{6} M_{\odot}\) and \(R_{0} \sim 8.3 \text{kpc}\). These quantities have been examined using a Newtonian approach; namely, a Keplerian central potential was assumed, in which the BH mass is treated as being concentrated in a point-like object \cite{SCH1,SCH2}.

The foundational study  \cite{l11} was established through the kinematic frequency variation, which is not straightforwardly derived and has subsequently been utilize across a broad spectrum of BH models such as Kerr-Newman BHs in de Sitter geometry \cite{l13A}, regular BHs in a spherically symmetric background \cite{l14}, and higher-dimensional BHs \cite{l15}.  
As an application of this framework, the Kerr BH parameters $M$ and $a$, together with the radius of the detector’s orbit $r_d$, can be determined from the redshift and blueshift of photons that propagate along null geodesics and are emitted by massive bodies orbiting the BH along arbitrary trajectories, such as the recently observed closed stellar orbits around the center of our Galaxy \cite{l11,l12,l16}.
In the last few decades, it has been demonstrated that a similar methodology can be applied to derive Hubble expansion from Kerr–dS geometry, using a novel discrete general relativistic method to estimate the late-time Hubble rate \cite{l13,l16A,l16B,l16C}.

{The radial component has been considered to develoment of BH in bumblebee field gravity: {$b_r(r)=|b| e^{\rho(r)}$}. The influence of the bumblebee field is encoded in the metric through the Lorentz-violating parameter {$\ell=\xi b^{2}$}, while the standard Schwarzschild BH solution is recovered in the Lorentz-invariant limit, $\ell\rightarrow0$.  Since the background vector {\(b_{\mu}\)} is assumed to be purely radial and spacelike, its associated field-strength tensor vanishes identically, $B_{\mu\nu}=0$. At the background level, it introduces the Lorentz-violating distribution, which modifies the spacetime geometry and leads to observable consequences.
However, most astrophysical BHs are embedded in a plasma medium, where the photon frequency is influenced by plasma effects \cite{l17,l18,l19}. Consequently, it is essential to analyze the bumblebee AdS BH within the general relativistic framework, and it is equally significant to determine the characteristics of plasma on the frequency shifts}.

The effects of bumblebee fields in cosmology have been investigated through dynamical equations characterizing the evolution of the Universe and its features. In Ref. \cite{l20,BBH1,BBH2}, a bumblebee field was used to produce cosmological anisotropies, acting as a source of anisotropies and introducing a distinguished axis. This leads to a contribution to cosmic anisotropies originating from Lorentz symmetry violation. The temperature of the bumblebee AdS BH was also investigated and remains positive throughout the parameter space considered.  Moreover, it was indicated that AdS BHs exhibit a phase transition with critical points that depend on the Lorentz-violating factor. In particular, this transition criticality decreases (increases) as $|l|$ increases when $l>0$ ($-1<l<0$). However, the region of local stability increases with increasing $l$ for $l>0$.

In recent years, increasing attention has been devoted to understanding BHs in realistic astrophysical environments, where realistic BH images are typically illuminated by electromagnetic radiation from nearby accretion flows \cite{zz1,zz2,MYAJ0,MYAJ1,MYAJ2,MYAJ3,MYAJ4}. However, our focus here is solely on the influence of the BH spacetime on photon propagation. Therefore, a particular prediction of this theory was presented, modeling an accretion disk, we adopt a celestial sphere as the background light source. We partition the celestial sphere into four quadrants, each assigned a distinct color (green, blue, red, and yellow) \cite{zz3,zz4,zz5,zz6,zz7,zz8}.

This work is structured as: In next section, we concisely discuss the  background solution of the bumblebee gravity theory. In \textbf{Section III}, we examine the geodesic motion of BHs and formulate the nonvanishing components of geodesic massive particles as metric functions. Subsequently, we proceed to determine the light-bending parameters as functions of the metric quantities.  
In \textbf{Section IV}, we derive a general mathematical formulation for the total observed frequency shift of radiated photons by the test particle. 
In \textbf{Section V}, we investigate the impacts of specific parameters of bumblebee AdS on frequency shift, demonstrating an application of our results. 
In \textbf{Section VI}, we present the shadow cast of BH using the backward ray-tracing process implemented in the Arcmancer code.
A summary of the main findings and conclusions is given in \textbf{Section VII}.

\section{Background of the black hole solutions in bumblebee gravity theory}
\subsection{Bumblebee AdS black holes}

The action of bumblebee gravity with a cosmological constant is formulated as
 \cite{BBH1, BBH2}
\begin{eqnarray}\nonumber
\mathcal{I}&=&\int_{\partial \mathcal{M}} d^4 x \sqrt{-g}\bigg[\frac{R-2 \Lambda}{2 \kappa}+\frac{\xi\left(B^\mu B^\nu R_{\mu \nu}\right)}{2 \kappa}-\frac{B_{\mu \nu} B^{\mu \nu}}{4}\\ \label{B1}
&-& V\left(B^\mu B_\mu \pm b^2\right)+\mathcal{L}_{\text {matter }}\bigg].
\end{eqnarray}
Here, \(\kappa=\frac{8\pi G}{c^4}\), while \(\Lambda\) denotes the cosmological constant and the parameter \(\xi\) represents the coupling constant which characterizes the nonminimal interaction between the bumblebee field and the Ricci tensor. Moreover, the antisymmetric tensor
\begin{equation}
B_{\mu\nu}=\partial_\mu B_\nu-\partial_\nu B_\mu,
\end{equation}
defines the field strength associated with the bumblebee field. The term \(\mathcal{L}_{\mathrm{matter}}\) provides the matter sector together with interaction terms involving the field \(B_\mu\). 
Varying the action in Eq.~(\ref{B1}) with respect to the metric
\(g_{\mu\nu}\), while treating \(B_\mu\) as a nondynamical field, yields the gravitational field equations
\cite{BBH1,BBH2}
\begin{equation}\label{B2}
G_{\mu \nu}+\Lambda g_{\mu \nu}=\kappa\left(T_{\mu \nu}^B+T_{\mu \nu}^M\right),
\end{equation}
where $G_{\mu \nu}$ represents the Einstein tensor, and $T_{\mu \nu}^B$ denotes the energy-momentum tensor associated with the bumblebee field, expressed as follows \cite{BBH1,BBH2}
\begin{eqnarray}\nonumber
T_{\mu \nu}^B&=& B_\mu^\alpha B_{\nu \alpha}+2 V^{\prime} B_\mu B_\nu-(V+\frac{1}{4} B_{\alpha \beta} B^{\alpha \beta}) g_{\mu \nu}\\ \nonumber
&+&\frac{\xi}{2 \kappa}[B^\alpha B^\beta R_{\alpha \beta} g_{\mu \nu}+\nabla_\alpha \nabla_\mu(B^\alpha B_\nu)+\nabla_\alpha \nabla_\nu\\ \nonumber
&&(B^\alpha B_\mu) 
-2 B_\mu B^\alpha R_{\alpha \nu}-2 B_\nu B^\alpha R_{\alpha \mu}\\ \label{B3}
&-&\nabla^2(B_\mu B_\nu)-g_{\mu \nu} \nabla_\alpha \nabla_\beta(B^\alpha B^\beta)]. 
\end{eqnarray}
Here, the operator denotes differentiation with respect to the argument of the potential. In the following, we adopt the natural units \(G=c=1\). Moreover, an additional equation of motion for
\(B^\mu\) is obtained by varying the action in Eq.~(\ref{B1}) with respect to the bumblebee field as
\begin{equation}\label{B4}
\nabla_\mu B^{\mu \nu}=2\left(V^{\prime} B^\nu-\frac{\xi}{2 \kappa} B_\mu R^{\mu \nu}\right).
\end{equation}
A four-dimensional static spacetime is given by
\begin{equation} \label{B5}
d s^2=-f(r) d t^2+ \frac{(1+l)}{f(r)}d r^2+g_{\theta \theta}d \theta^2+g_{\varphi \varphi}d \varphi^2.
\end{equation}
Here, the metric elements are $g_{tt}=f(r)$, $g_{rr}=g(r)={(1+l)}/f(r)$, $g_{\theta \theta}=r^2$ as general spacetimes of coordinate, radial coordinate and $g_{\varphi \varphi}=r^2 \sin^2 \theta$, respectively. Subsequently, we omit the arguments $r$ and $\theta$ from the spacetimes for notational convenience. Furthermore, the bumblebee field parameter is formulated in that metric by $l=\xi b^2$. It is noteworthy that $b=\sqrt{b_\mu b^\mu}$, which is defined as follows:
\begin{equation}\label{B6}
B_\mu=b_\mu=\left(0, b_r(r), 0,0\right),
\end{equation}
such that $b_\mu$ is the underlying field, and it is a spacelike vector aligned purely radially. Also, to preserve the signature of the aforementioned spacetime Eq.~(\ref{B5}), we have to enforce the condition $l > -1$. Using Eqs.~(\ref{B2})–(\ref{B6}), the lapse function is expressed as
 \cite{BBH2}
\begin{equation}\label{B77}
f(r)=1-\frac{2 m_0}{r}-(1+l) \frac{\Lambda}{3} r^2.
\end{equation}
Here, $\Lambda = \frac{\kappa \lambda}{\xi}$ is recognized as an effective cosmological constant. Additionally, $\lambda$ denotes a Lagrange multiplier field.
Our findings reveal that the Lorentz-violating parameter $\ell=\xi b^{2}$ and the effective cosmological constant $\Lambda$ impose fundamental constraints on the location of stable particle orbits around BHs, thereby establishing a direct connection between theoretical predictions and potentially observable astrophysical phenomena.

\subsection{Bound and Divergence Points}

The roots of the temperature can be examined to distinguish between the physical and non-physical features of bumblebee Schwarzschild-(A)dS BHs.
A positive temperature corresponds to a physically admissible solution, whereas a negative temperature indicates a non-physical configuration.
After determining the thermodynamic behavior of the bumblebee Schwarzschild-(A)dS BH, it is important to investigate the phase-transition points and the bound points of the heat capacity. 
The heat capacity (specific heat) exhibits two types of roots and one divergence point. The zeros of the heat capacity, or equivalently the temperature, represent the bound points, whereas the divergence of the heat capacity corresponds to a phase-transition point.
Moreover, the BH solution remains physical within the region bounded by the small and large roots. Based on the above discussion, the roots and divergence points of the heat capacity can be further analyzed as follows:
\begin{equation}
C=\frac{T}{\partial^2M/\partial S^2}.
\end{equation}
Using $S=\pi r_+^2$, the heat capacity can be expressed as 
\begin{equation}
C=-2\pi r_+^2\frac{1-(1+l)\Lambda r_+^2}{1+(1+l)\Lambda r_+^2}.
\end{equation}
\textbf{Bound points:}\\
The bound points associated with to the zeros of the heat capacity are obtained as
\begin{equation} \label{bp1}
r_{+,BP}^{(1)}=\frac{1}{\sqrt{(1+l)\Lambda}},
\end{equation}
and 
\begin{equation} \label{bp2}
r_{+,BP}^{(2)}=-\frac{1}{\sqrt{(1+l)\Lambda}}.
\end{equation}
To investigate the bound point to be real, these parameters must satisfy $(1+l)\Lambda>0$.
So the bound point denotes the extremal boundary of the thermodynamics physical region.
\textbf{Divergence point:}\\
The divergence of the heat capacity is obtained when its denominator is set to zero
\begin{equation}
r_{+,DP}^{(1)}=+\frac{1}{\sqrt{-(1+l)\Lambda}},
\end{equation}
and
\begin{equation}
r_{+,DP}^{(2)}=-\frac{1}{\sqrt{-(1+l)\Lambda}}.
\end{equation}
In this case, the  physically relevant positive divergence point one requires $(1+l)\Lambda<0$, and  the divergence radius becomes
\begin{equation}
r_{+,DP}=\frac{1}{\sqrt{(1+l)|\Lambda|}}.
\end{equation}
For the physical interpretation, we consider the AdS sector and bound-point condition such as: $\Lambda<0, 1+l>0$ and  $(1+l)\Lambda>0$, respectively.
However, the heat capacity changes sign at horizon radius as 
\begin{equation}
0<r_+<r_{+,DP}, \quad C<0,
\end{equation}
which provides information that small AdS BHs are thermodynamically unstable, whereas
\begin{equation}
r_+>r_{+,DP}, \quad C>0,
\end{equation}
corresponds to the locally stable large BH. So, we have examined that the bumblebee Schwarzschild-(A)dS model exhibits two bound roots and one divergence point.
\[
\begin{array}{c|c}
\hline
\text{Sector} & \text{Characteristic point} \\
\hline
(1+l)\Lambda>0
& r_{+,BP}=
\dfrac{1}{\sqrt{(1+l)\Lambda}}
\\[8pt]
(1+l)\Lambda<0
&
r_{+,DP}=
\dfrac{1}{\sqrt{-(1+l)\Lambda}}
\\
\hline
\end{array}
\]
Hence, in the case of $\Lambda<0, \quad l>-1$, the physically relevant characteristic is the  heat-capacity (single) divergence  instead of small and large bound-point roots.

\begin{figure}
\includegraphics[width=16pc]{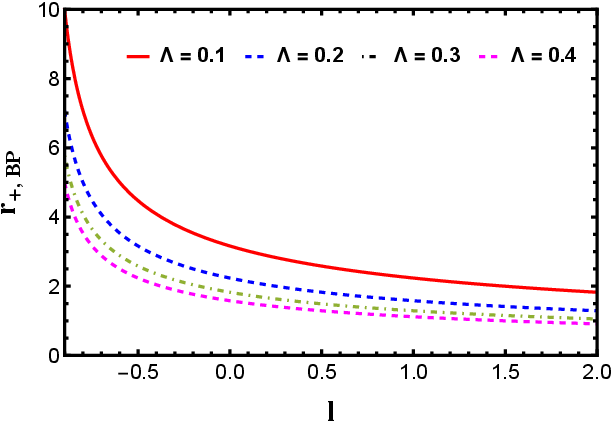}
\caption{\label{BP1} {$r_+$,BP versus $l$ for Eq. (\ref{bp1}) diagram for different values of  $\Lambda$ in the bumblebee AdS BH.}}
\end{figure}
In addition, we plot Fig.~\ref{BP1} to examine the influence of the cosmological constant \(\Lambda\) on the bound points. We observe that the bound points decrease as \(\Lambda\) increases. It is worth noting that increasing \(\Lambda\) enlarges the physical and thermodynamically stable region. Consequently, the stability of large BHs is enhanced. The figure also clearly indicates that the divergences of the heat capacity disappear for sufficiently large values of \(\Lambda\). Interestingly, a similar phase-space analysis can be performed using Eqs.~(\ref{bp1}) and (\ref{bp2}), with the parameter space restricted to \(\Lambda>0\).

\subsection{Perturbation theory analysis}

To study the linear stability of the background solution, we perturb both the metric and the bumblebee vector field by introducing
$\epsilon h_{\mu\nu}$ and $\epsilon\,\delta b_{\mu}$, respectively, with $\epsilon\ll1$. The corresponding expansions of the background metric $g_{\mu\nu}$ and vector field $b_{\mu}$ take the form \cite{PT1} 
\begin{equation}\label{eq33}
g_{\mu\nu}\rightarrow g_{\mu\nu}+\epsilon h_{\mu\nu},
\quad b_{\mu}\rightarrow b_{\mu}+\epsilon \delta b_{\mu}.
\end{equation}
For a spherically symmetric background spacetime, the linear perturbations can be classified into odd- and even-parity sectors according to their behavior under parity transformations. In particular, the metric perturbation components $h_{tt}$, $h_{tr}$, and $h_{rr}$ transform as scalar fields under rotations on the two-dimensional sphere $(\theta,\varphi)$. Consequently, an arbitrary scalar function $\Psi(t,r,\theta,\varphi)$ admits the harmonic decomposition
\begin{equation}\label{eq33A}
\Psi(t,r,\theta,\varphi)=\sum_{\ell,m}\Psi_{\ell m}(t,r)Y_{\ell m}(\theta,\varphi),
\end{equation}
where $Y_{\ell m}(\theta,\varphi)$ are the standard spherical harmonic functions.
\begin{equation}\label{eq34}
\Psi(t,r,\theta,\varphi)=\sum_{\ell,m}\Psi_{\ell m}(t,r)
Y_{\ell m}(\theta,\varphi).
\end{equation}
The parity properties of the spherical harmonics are encoded in the
transformation
\begin{equation}\label{eq34A}
Y_{\ell m}(\pi-\theta,\varphi+\pi) =
(-1)^{\ell}Y_{\ell m}(\theta,\varphi),
\end{equation}
Applying the parity transformation on the unit two-sphere yields
\begin{equation}\label{eq34B}
\Psi(t,r,\pi-\theta,\varphi+\pi)=(-1)^{\ell} \Psi(t,r,\theta,\varphi).
\end{equation}
The above transformation characterizes the even-parity sector. In contrast, for $a,b=\theta,\varphi$, the perturbation components $h_{ta}$ and $h_{ra}$ transform as vectors, whereas $h_{ab}$ transforms as a rank-two tensor on the two-sphere. The corresponding vector and tensor harmonics can be classified according to their parity: modes transforming as $(-1)^{\ell}$ belong to the even-parity sector, while those transforming as $(-1)^{\ell+1}$ constitute the odd-parity sector \cite{PT2}.  In our work, the bumblebee field is taken to be purely radial and frozen at its vacuum expectation value, with constant norm $B_{\mu}B^{\mu}=b^{2}$. 
We have expanded the motivation and relevant background underlying the present analysis to better clarify the rationale of our study and the significance of the adopted framework.

\subsection{Existence of a ghost mode}
{ The ghost and strong-coupling issues discussed in the literature concern the full perturbative dynamics of certain bumblebee models. Nilsson \cite{GH1} demonstrated that nonminimally coupled bumblebee gravity generally propagates a ghost degree of freedom unless suitable degeneracy conditions are imposed, whereas the minimally coupled theory suffers from strong-coupling pathologies. More recently, these proceedings reviewed the latest developments in a class of vector--tensor theories, commonly known as bumblebee models, in which local Lorentz and diffeomorphism symmetries are spontaneously broken. Using cosmological perturbation theory on a FLRW background, it was shown that the inclusion of a nonminimal coupling to gravity generically introduces a ghost mode unless the required degeneracy conditions are satisfied. When these conditions hold, the theory reduces to a particular subclass of generalized Proca theory, with the bumblebee potential uniquely determined by the background field equations \cite{GH2}.\\
\textbf{The model}:
We begin with the most general bumblebee action incorporating a nonminimal coupling to gravity, given by
\begin{equation}
\begin{aligned}
\mathcal{L}_{B}\sim &\frac{M_{\mathrm{Pl}}^{2}}{2}R
-\frac{\gamma}{4}B_{\mu\nu}B^{\mu\nu}-V(B^{2})+\eta B^{2}\nabla_{\mu}B^{\mu}\\
&+\xi B^{\mu}B^{\nu}R_{\mu\nu}+\sigma B^{2}R+\varsigma\left(\nabla_{\mu}B^{\mu}\right)^{2}+\upsilon \nabla_{\mu}B^{\mu}R ,
\end{aligned}
\label{gh1}
\end{equation}
It is worth emphasizing that, unlike in most of the existing bumblebee literature, the action is formulated such that all coupling constants, \(\{\gamma,\eta,\xi,\sigma,\varsigma,\upsilon\}\),
are dimensionless. Furthermore, \(M_{\mathrm{Pl}}\) denotes the Planck mass, while \(V_{B}\) represents the potential of the bumblebee field \(B^{\mu}\), whose nonvanishing vacuum expectation value spontaneously breaks diffeomorphism invariance and, consequently, local Lorentz symmetry. Since all operators appearing in the action are marginal, the only dimensionful scale in the theory is generated through the bumblebee potential.
We study cosmological perturbations of the model~(\ref{gh1}) around exact dS, and we decompose the bumblebee field as
\begin{equation}\label{eqgh6}
B_{\mu}\rightarrow \left\{\bar{B}_{0}+\epsilon\,\delta B_{0},\epsilon \partial_{i}\delta B_{s}+\epsilon\,\delta B_{i}^{(T)}\right\},
\end{equation}
The perturbative decomposition contains two scalar modes together with one transverse (divergence-free) vector mode; however, not all of these constitute propagating dynamical degrees of freedom. Accordingly, we express the metric in the Arnowitt--Deser--Misner (ADM) form as
\begin{equation}\label{eqgh7}
ds^{2}=-N(t)^{2}dt^{2} +a(t)^{2}\gamma_{ij}\left(dx^{i}+N^{i}dt\right)\left(dx^{j}+N^{j}dt\right),
\end{equation}
which is perturbed in spatially-flat gauge as
\begin{equation} \label{eqgh8}
N(t)=1+\epsilon\alpha,\quad N^{i}=\frac{\epsilon}{a}
\left(\mathcal{B}^{i}+\partial^{i}\beta\right),
\quad \gamma_{ij}=\delta_{ij}+\epsilon h_{ij},
\end{equation}
The perturbation sector consists of two scalar modes, one transverse (divergence-free) vector mode, and two symmetric, traceless tensor modes. We then expand the action to quadratic order in the linear perturbations of the scalar fields \((\alpha,\beta,\delta B_{0},\delta B_{s})\), the vector fields \(\left(\mathcal{B}^{i},(\delta B^{\perp})^{i}\right)\), and the tensor field \((\gamma_{ij})\). Not all of these perturbative degrees of freedom correspond to independent propagating dynamical modes. In what follows, we focus on the dynamics of the tensor and scalar sectors.\\
\textbf{Scalar perturbations}:
After going to Fourier space and integrating by parts, we straightforwardly find that $\beta$ is non-dynamical and we integrate it out. Due to the presence of the coupling $\upsilon$, the quadratic action contains $\ddot{\tilde{\alpha}}$ and $\delta\ddot{B}_{0}$, which we eliminate by the variable redefinition
$\tilde{\alpha}\equiv\alpha-\frac{\delta B_{0}}{\bar{B}_{0}}$
and the introduction $Q\equiv\dot{\tilde{\alpha}}$, after which we can write, to quadratic order in scalar perturbation,
\begin{equation}
\mathcal{L}_{\rm tot}^{(2)}\equiv\mathcal{L}_{2}^{(2)}+\mathcal{L}_{3}^{(2)}
+\lambda\left(Q \dot{\tilde{\alpha}}\right)=\mathcal{L}_{2}^{(2)}+\mathcal{L}_{3}^{(2)}+\lambda Q+\dot{\lambda}\tilde{\alpha}+\mathrm{T.D.},
\label{eqgh11}
\end{equation}
where \(\lambda\) is a Lagrange multiplier that enforces the constraint \(Q=\dot{\tilde{\alpha}}\). Eliminating \(\tilde{\alpha}\) through its equation of motion, the action can be expressed solely in terms of the variables \(\{\dot{\lambda},\lambda,\dot{Q},Q\}\). The resulting Lagrangian then takes the form
\begin{equation} \label{eqgh12}
\mathcal{L}^{(2)}=\frac{a^{3}}{2}\int d^{3}xdt
\left[\dot{\mathcal{V}}_{4}^{\dagger}\mathbf{K}_{4}
\dot{\mathcal{V}}_{4}+k\dot{\mathcal{V}}_{4}^{\dagger}
\mathbf{F}_{4}\mathcal{V}_{4}-\mathcal{V}_{4}^{\dagger}
\mathbf{X}_{4}\mathcal{V}_{4}\right],
\end{equation}
where \(\mathbf{K}_{4}\) represents the kinetic (Hessian) matrix,
\(\mathbf{F}_{4}\) denotes the friction matrix, and
\(\mathbf{X}_{4}\) comprises the gradient and mass matrices.
We further define the field vector
\[\mathcal{V}_{4}\equiv (\lambda,Q,\delta B_{0},\delta B_{s}).
\]
Imposing this condition on the quadratic action~(\ref{eqgh11}), we find that the Lagrange multiplier \(\lambda\) can be integrated out, yielding
\begin{equation}
\mathcal{L}_{2}^{(2)}=
\frac{1}{2}a^{3}\left(\dot{\nu}_{3}^{\dagger}\mathbf{K}_{3}\dot{\nu}_{3}
+\dot{\nu}_{3}^{\dagger}\mathbf{N}_{3}\nu_{3}-\nu_{3}^{\dagger}\mathbf{X}_{3}\nu_{3}\right),
\end{equation}
where \[\nu_{3}^{\dagger}\equiv(\alpha,\delta B_{0},\delta B_{s}),
\]
and the kinetic matrix is given by
\begin{equation}
\mathbf{K}_{3}=
\begin{pmatrix}
K\bar{B}_{0}^{2} & -K\bar{B}_{0} & 0 \\
-K\bar{B}_{0} & K & 0 \\
0 & 0 & \gamma\dfrac{k^{2}}{a^{2}}
\end{pmatrix},
\end{equation}
with
\begin{equation} \label{eqgh13}
K\equiv-\frac{2}{s}(\xi+2\sigma)(\xi+2\sigma+2\varsigma).
\end{equation}
The kinetic matrix satisfies \(\mathrm{rank}(\mathbf{K}_{3})=2\), indicating that the theory generically propagates two scalar degrees of freedom. From above equations, we observe that there exist two nontrivial choices for which the rank is reduced to unity. We first consider
We then compute the quadratic action for the scalar perturbations. In this form, both \(\beta\) and \(\alpha\) can be integrated out, leaving a rank-two kinetic matrix of the form
$\mathbf{K}_{2}=\operatorname{diag}\left(18(\xi+2\sigma)^{2}\frac{H^{2}\bar{B}_{0}^{2}}{D},\gamma\frac{k^{2}}{a^{2}}
\right)$, where \(D\) denotes a lengthy expression. Since the rank of the kinetic matrix remains greater than unity, further degeneracy conditions must be imposed to obtain a healthy theory. As follows from Eq.~(\ref{eqgh13}), there are two possible branches:
\begin{align}
\xi+2\sigma &=0, \quad \text{and} \quad \xi+2\sigma+2\varsigma\neq0,
\label{eqgh17}
\\
\xi+2\sigma &\neq0,
\quad \text{and}\quad \xi+2\sigma+2\varsigma=0.
\label{eq18}
\end{align}
By considering both branches, we find that the necessary and sufficient condition for the theory to propagate a single scalar degree of freedom is
\begin{equation} \label{eqgh19}
\boxed{\xi+2\sigma=0 \quad\text{and} \quad \varsigma=0.}
\end{equation}
Therefore, we conclude that the above condition is the unique choice that yields the correct number of propagating degrees of freedom. Consequently, limits such a \(\{\sigma=0,\xi\neq0\}\) cannot be consistently realized. In other words: \textit{\textbf{The bumblebee model (\ref{gh1})  has a ghost instability unless the degeneracy condition (\ref{eqgh19}) is imposed}.}
In particular, we emphasize that our objective is to investigate the geodesic structure of massive particle motion, examine the possible CBV, and explore the corresponding ray-tracing signatures in the bumblebee BH spacetime}. 

\section{Motion of Timelike and Null Particles along Geodesics}

The following dynamical relation that governs the motion of orbiting particles is given by
\cite{RD12,RD13,RD15,RD16,RS1}
\begin{equation}\label{B7}
{d^2s}/{d \lambda^2}=\kappa=g_{\mu \nu} \dot{x}^\mu \dot{x}^\nu.
\end{equation}
In this scenario, $\kappa = -1,0$  provide the information of timelike and null particles, respectively, therefore $\dot{x}^\mu = \frac{d x^\mu}{d \lambda}$ represents the derivative with respect to $\lambda$, which serves as the affine parameter along the photon trajectory, and this corresponds to the proper time ($\lambda = \tau$) for massive particles. So, we assume that the timelike Killing vector field as
\begin{equation}\label{B8}
\Phi^\mu=(1,0,0,0),
\end{equation}
and rotational case
\begin{equation}\label{B8A}
\varrho ^\mu  =(0,0,0,1).
\end{equation}
These functions reflect the symmetries of the metric element, and this relation can be utilized to derive the conserved quantities associated with geodesic motion.
Without loss of generality, we restrict the BH trajectory to the equatorial plane at $\theta = \pi / 2$.
The oblique geodesic trajectories of massive particles as observed from the Earth, in this case, a rotational transformation of the reference frame using the Euler method angles can be applied \cite{RD23,RD24,RD25,RD26}.

\section{Geodesic Structure of Massive Particle Motion}

The geodesic equation (\ref{B7}) of motion for particles is given by  \cite{RD12,RD13,RD16,RS1}
\begin{equation}\label{B9}
g_{\mu \nu} \Upsilon^\mu \Upsilon^\nu=-1,
\end{equation}
where $\Upsilon^\mu$ denotes the 4-velocity vector of an object. Alternatively, the invariant quantities corresponding to the Killing vector field (\ref{B8}) implies 
\begin{equation}\label{B10}
E=\frac{\bar{E}}{m}=-\Phi_\mu \Upsilon^\mu=-f(r) \Upsilon^t,
\end{equation}
and 
\begin{equation}\label{B10A}
L=\frac{\bar{L}}{m}=\psi_\mu \Upsilon^\mu=g_{\varphi \varphi} U^{\varphi},
\end{equation}
where $E = \tilde{E}/m$ and $ L = \tilde{L}/m$ represent the energy coupled with angular momentum per unit mass, with  $\tilde{E}$ and $\tilde{L}$ being the total quantities. 
We will study the particle’s trajectory constrained to the equatorial plane where $\Upsilon^\theta=0$, which is given in Eq. (\ref{B9}), this provides the form of an energy balance relation. Typically, the first part of the above relation corresponds to the kinetic energy (KE) of a test particle influenced by an effective potential as
\begin{eqnarray} \nonumber
V_{eff}&=&- f(r)\bigg(1+\frac{L^2}{g_{\varphi \varphi}}\bigg)=\bigg(\frac{\Lambda  (l+1) r^2}{3}+\frac{2 {m_0}}{r}-1\bigg) \\ \label{B11}
&&\bigg(\frac{L^2 \csc ^2(\theta )}{r^2}+1\bigg).
\end{eqnarray}
Taking $\theta = \pi / 2$ enables the formulation of certain expressions in a simplified manner without any loss of generality. From Eqs. (\ref{B10}) and (\ref{B10A}), the following criteria regulate the motion of circular paths followed by massive particles we have \cite{RD15,RD16,RS1}
\begin{equation}\label{B11A}
-1-\frac{E^2}{f(r)}-\frac{L^2}{g_{\varphi \varphi}}=V_r(r)==0, 
\end{equation}
and
\begin{equation}\label{B11Aa}
\frac{E^2}{f(r)^2} f^{\prime}(r)+\frac{L^2}{g_{\varphi \varphi}^2} g_{\varphi \varphi}^{\prime}=V_r^{\prime}(r)=0,
\end{equation}
we obtain 
\begin{equation}\label{B12}
V_r(r)=\frac{3 E^2 r}{\Lambda  (l+1) r^3+6 {m_0}-3 r}-\frac{L^2 }{r^2}-1,
\end{equation}
\begin{equation}\label{B13}
V_r^{\prime}(r)=\frac{6 E^2 \left(3 {m_0}-\Lambda  (l+1) r^3\right)}{\left(\Lambda  (l+1) r^3+6 {m_0}-3 r\right)^2}+\frac{2 L^2}{r^3},
\end{equation}
and 
\begin{equation}\label{B14}
E^2=g_{\varphi \varphi}^{\prime}\left(\frac{f(r)^2}{g_{\varphi \varphi} f^{\prime}(r)-f(r) g_{\varphi \varphi}^{\prime}}\right), 
\end{equation}
\begin{equation}\label{B14A}
L^2=f^{\prime}(r)\left(\frac{g_{\varphi \varphi}^2}{f(r) g_{\varphi \varphi}^{\prime}-g_{\varphi \varphi} f^{\prime}(r)}\right).
\end{equation}
Using the above expressions, the results obtained as 
\begin{equation}\label{B14C}
E^2=-\frac{\left(\Lambda  (l+1) r^3+6 {m_0}-3 r\right)^2}{9 r (r-3 {m_0})}, 
\quad
L^2=\frac{r^5}{r-3 {m_0}}.
\end{equation}
Here, the frame-dragging effects experienced by test particles moving in non-equatorial regions of a spherically symmetric spacetime are expressed in terms of generalized hypergeometric functions of the Appell and Lauricella types. Based on Eqs.~(\ref{B10}) and (\ref{B10A}), the ($t$)- and ($\varphi$)-components can be formulated as
\begin{equation}\label{B15C}
U_e^t=-\frac{E}{f(r)}=\frac{\Lambda  (l+1) r^3+6 {m_0}-3 r}{9 {m_0}-3 r},
\end{equation}
\begin{equation}\label{B15D}
U_e^{\varphi}=\frac{L}{g_{\varphi \varphi}}=\frac{r^3}{r-3 {m_0}}.
\end{equation}
For the persistence of circular trajectories, it is necessary that the second derivative of $V_r(r)$ satisfies the required stability criterion as follows:
\begin{eqnarray} \nonumber 
V_r^{\prime \prime}(r)&=& -\frac{2 E^2}{f(r)^3}\left(f^{\prime}(r)\right)^2+\frac{E^2}{f(r)^2} f(r)^{\prime \prime} 
-\frac{2 L^2}{g_{\varphi \varphi}^3}\left(g_{\varphi \varphi}^{\prime}\right)^2
\\ \label{B16} 
&+&\frac{L^2}{g_{\varphi \varphi}^2} g_{\varphi \varphi}^{\prime \prime} \leq 0.
\end{eqnarray}
We provide derivations of elegant and compact formulations for the fixed parameters of motion in these orbits.
\begin{figure}
\includegraphics[width=16pc]{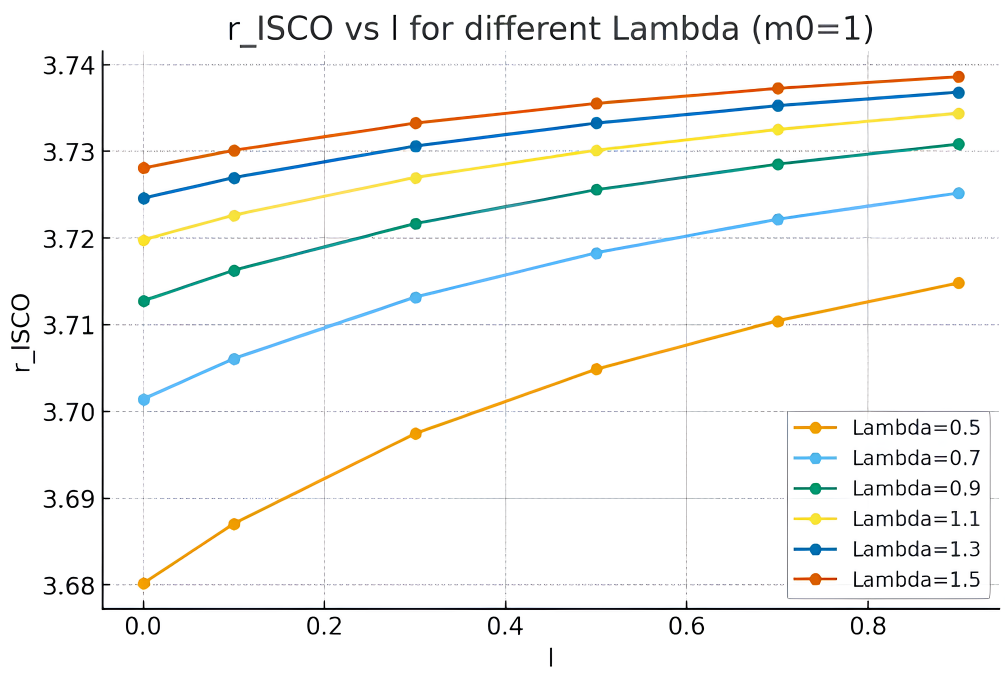}
\caption{\label{riscobumblebee} {Variation of \( r_{\mathrm{ISCO}} \) with \( l \) for different values of \(\Lambda\) in the bumblebee AdS BH (\(m_0 = 1\)).}}
\end{figure}
Furthermore, by combining these relations, one can obtain closed-form expressions for the redshift/blueshift of photons emitted by test particles in non-equatorial with fixed-radius orbits around BH. From Eq. (\ref{B16}), we get 
\begin{equation}\label{B16C}
V_r^{\prime \prime}(r)=\frac{6 {m_0} \left(5 \Lambda  (l+1) r^3+r\right)-8 \Lambda  (l+1) r^4-36 {m_0}^2}{r^2 (r-3 {m_0}) \left(\Lambda  (l+1) r^3+6 {m_0}-3 r\right)}.
\end{equation}

\subsection{ISCO formulation}

To evaluate expression (\ref{B16C}), we have
\begin{equation}\label{cht1}
V_r''(r)=\frac{N(r)}{D(r)}.
\end{equation}
In this regime, $V_r''$ assumes a sign indicating that the effective potential behaves as a local minimum under the adopted convention.
In practice, the sign of the inequality can be identified by evaluating $V_r''$ at a reference radius corresponding to stable orbits.  We employ different $V$-conventions to establish the stability conditions such that $r_{\rm test}\gg m_0$ where Keplerian orbits at large distances are expected to be stable. Hence, the physical behavior is determined by $V_r''>0$. Use the min–max bounds form as \cite{S1,S2}
$$
\frac{\min N}{\max D}\le V_r''(r)\le \frac{\max N}{\min D}.
$$
We notice that if $\dfrac{\min N}{\max D}>0$, then $V_r''>0$ throughout the interval, providing the stability for all $r>r_h$. Conversely, if $V_r''(r \gg m_0)<0$ for $r>r_h$, the configuration becomes unstable.
Consequently, the ISCO is determined by the marginal stability condition by using the Eqs.~(\ref{B14}), (\ref{B14A}) and (\ref{B16}) , we obtain
\begin{equation} \label{Bisco}
\frac{4(1+l)\Lambda}{3}r_{\rm ISCO}^4-5(1+l)\Lambda m_0 r_{\rm ISCO}^3-m_0 r_{\rm ISCO}+6m_0^2=0.
\end{equation}
In the limiting case $\Lambda=0$, one immediately recovers the Schwarzschild result
$r_{\rm ISCO}=6m_0$.
In Table \ref{T1}, we analyze ISCO for the bumblebee AdS BH, which shows that increasing the cosmological constant $\Lambda$ deepens the effective potential and shifts stable orbits outward. Further, it enlarges the ISCO radius compared to asymptotically flat cases.
\begin{figure}
\includegraphics[width=16pc]{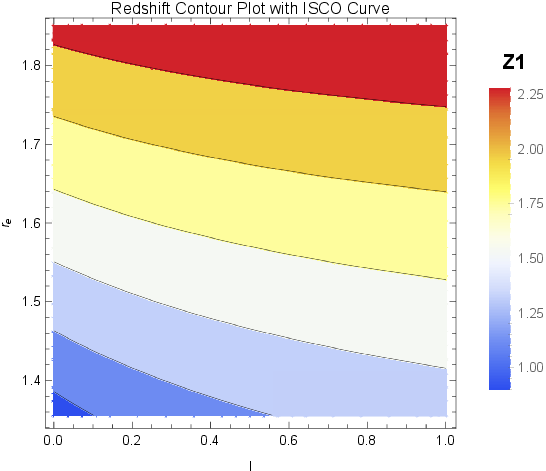}
\caption{\label{fig1} The redshift $z_{1}$ curves in bumblebee AdS BH with parameter $m_0=1$.}
\end{figure}
For larger $l$, the ISCO radius differs significantly from the Schwarzschild–AdS background.  

{In Fig.~\ref{riscobumblebee}, the \( r_{\mathrm{ISCO}}\) for  BH is analyzed using the effective potential method. 
The event horizon radius is denoted by $r_h$, while the photon sphere radius is given by: $r_{ps}=\sqrt[3]{3}/\sqrt[3]{\Lambda+\Lambda l+3}$. The corresponding BH shadow radius is defined as $R_s=r_{ps}/\sqrt{f(r_h)}$. For a physically consistent BH geometry, the well-known inequality: $r_h \prec r_{ps} \prec R_s \prec r_{\mathrm{ISCO}}$ must be satisfied.
By imposing suitable energy conditions on the external matter fields, we derive the characteristic ISCO equation and show that this radius is bounded by $r_{\mathrm{ISCO}} \leq 6m_0$, where $m_0 = 1$ for the bumblebee AdS BH.
In the Schwarzschild background, the \(r_{\mathrm{ISCO}}\) saturates this bound (\( r_{\mathrm{ISCO}} = 6m_0\)). This behavior emphasizes the intricate interplay between BH geometry and external matter fields.} 
\begin{table}[h!]
\centering
\renewcommand{\arraystretch}{1.1} 
\setlength{\tabcolsep}{12pt}      
\begin{tabular}{|c|c|c|c|c|}
\hline
\textbf{$l$} &  \textbf{$r_{h}$} & \textbf{$r_{ph}$} & \textbf{$r_{sh}$} & \textbf{ISCO} \\
\hline
0.88 & 1.1627 & 1.2886 & 2.2348 & 4.7840 \\
0.89 & 1.1644 & 1.2854 & 2.2775 & 4.8028 \\
0.90 & 1.1662 & 1.2822 & 2.3236 & 4.8224 \\ 
0.91 & 1.1680 & 1.2790 & 2.3736 & 4.8428 \\
0.92 & 1.1699 & 1.2757 & 2.4281 & 4.8640 \\
\hline
\end{tabular}
\caption{Evaluation of $r_{\text{ISCO}}$ for different combinations $\Lambda=-0.85,-0.84,-0.83,-0.82,-0.81$ and $m_0=1$.}
\label{T1}
\end{table}

\begin{figure}
\includegraphics[width=16pc]{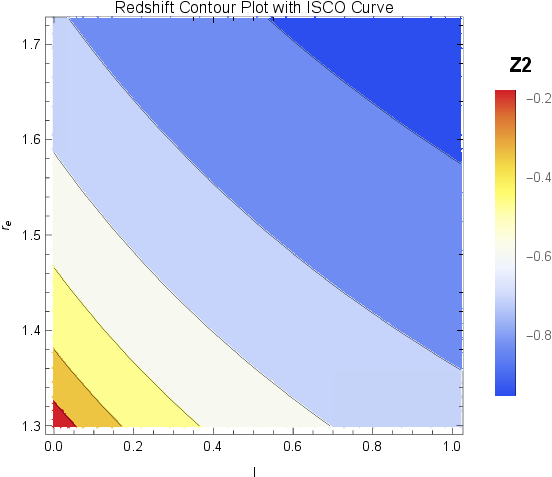}
\caption{\label{fig2} The redshift $z_{2}$ curves in bumblebee AdS BH with parameter $m_0=1$.}
\end{figure}

Blueshift occurs when the wavelength of light shortens, causing it to move toward the blue end of the spectrum, which leads to an object moving toward the observer, causing the light it emits to be compressed.
Consequently, our investigation focuses on the specified values \(m_0=1\) and the range \(0 \leq l \leq 1\) for the Lorentz-violating parameter of the bumblebee AdS BH. Figs.~\ref{fig1} and \ref{fig2} present density maps of the redshift and blueshift within the \(r_e\)–\(l\) plane for \(m_0=1\).
As evident from these figures, opposing effects can be examined through the redshift and blueshift profiles with respect to $r_e$ and $l$. As $r_e$ decreases, the magnitude of the frequency shift increases (decreases). This suggests that these parameters should compensate for the frequency shift at specific locations, indicated by the smooth gray contour lines, which correspond to the constant level surfaces of $z_{1,2}$.

\section {Geodesics of null particles}

In this section, we discuss the physical phenomenon of how light rays bend near BHs due to gravitational effects. For massless particles that travel along null geodesics, the equation of motion from Eq. (\ref{B7}) is  given by \cite{RD16,RS1}
\begin{equation}\label{B18}
g_{\mu \nu} k^\mu k^\nu=0,
\end{equation}
where \( k^\mu = \left(k^t, k^r, k^\theta, k^{\varphi}\right) \) denotes the four-momentum vector of the photons. So far, we propagate in regions outside the BH's event horizon, characterised by the specific forms of \( f(r) \), \( g_{rr}\), and \( g_{\varphi\varphi}\), respectively. By restricting the photon motion to the equatorial plane, the \(k^\theta\) component vanishes, and thus Eq.~(\ref{B18}) implies 
\begin{equation}\label{B19}
g_{r r}\left(k^r\right)^2+f(r)\left(k^t\right)^2=-g_{\varphi \varphi}\left(k^{\varphi}\right)^2.
\end{equation}
The symmetries inherent in the spacetime, as defined by the Killing vector fields in Eqs.~(\ref{B8}) and (\ref{B8A}), yield conserved quantities associated with geodesic motion, which follow as
\begin{equation}\label{B20}
E_\gamma=-\Phi_\mu k^\mu=-f(r) k^t,\quad
L_\gamma=\psi_\mu k^\mu=g_{\varphi \varphi} k^{\varphi}.
\end{equation}
Here, $L_\gamma$  and  $E_\gamma$ are the angular momentum and energy of the photons, respectively.
After that, we substitute these findings into the dynamical Eq.~(\ref{B18}), which describes the photon trajectories as
\begin{equation}\label{B21}
g_{r r}\left(k^r\right)^2+\frac{E_\gamma^2}{f(r)}+\frac{L_\gamma^2}{g_{\varphi \varphi}}=0 .
\end{equation}
Note that, at the radial coordinate corresponding to antipodal points, the redical component of the four-wave vector, \( k^r \), vanishes. Introducing the light deflection parameter in the following form
$
b_\gamma=\frac{L_\gamma}{E_\gamma}.
$ 
This expression can be formulated in terms of spacetime invariants as
\begin{equation}\label{B24}
b_{\gamma \pm}= \pm \sqrt{-\frac{g_{\varphi \varphi}}{f(r)}}.
\end{equation}
It is noteworthy that recent observational findings indicate that stellar trajectories are elliptical and do not reside within the equatorial plane. Consequently, a physically meaningful analysis requires evaluating the redshift and blueshift for \(\theta \neq \pi/2\).

\subsection{Frequency shift}

The photon frequency at specified position $x_p^\mu=\left(x^t, x^r\right.$, $\left.x^\theta, x^{\varphi}\right)\left.\right|_p$ reads \cite{RD15,RD16,RS1}
\begin{equation}\label{B24A}
\omega_p=-\left.\left(k_\mu \Upsilon^\mu\right)\right|_p,
\end{equation}
here the index $p$ denotes the photon's emission point $x_e^\mu$ or detection point $x_d^\mu$. Let us define the frequency shifts $\omega_p$  as
\begin{eqnarray}\label{B25}
1+z_{B H} & =&\frac{\omega_e}{\omega_d}\\ \nonumber
&=&\frac{\left.\left(E_\gamma \Upsilon^t-L_\gamma U^{\varphi}-g_{r r} \Upsilon^r k^r-g_{\theta \theta} \Upsilon^\theta k^\theta\right)\right|_e}{\left.\left(E_\gamma \Upsilon^t-L_\gamma U^{\varphi}-g_{r r} \Upsilon^r k^r-g_{\theta \theta} \Upsilon^\theta k^\theta\right)\right|_d}.
\end{eqnarray}
Furthermore, since the evolution of massive particle orbits is confined to the equatorial plane, this implies $\Upsilon^\theta = 0$, while the condition for circular trajectories requires $\Upsilon^r = 0$. Taking into account the observer at an asymptotically large radial distance from the particle, and the associated 4-velocity vector reduces to $U_d^\mu=(1,0,0,0)$ is given by
\begin{equation}\label{B25B}
z_{B H_{1,2}}+1=-b_{\gamma_{\mp}} U_e^{\varphi}+U_e^t,
\end{equation}
here the light deflection parameter is given in Eq. (\ref{B24}). It should be noted that $U_e^{\varphi}$ presented in Eq. (\ref{B25}), and also exhibits $\pm$ signs corresponding to Eqs. (\ref{B14}) and (\ref{B14A}), characterizing the rotation sense of massive particles ($+/-$ indicating counterclockwise/clockwise trajectories). If the choice between co-rotating and counter-rotating configurations is physically irrelevant, then we will proceed with the positive sign without loss of generality. In this scenario, the frequency shift of photon sources in circular orbits around a general spherically symmetric BH spacetime at radius $r=r_e$. From the above expression, $z_{BH_1}/z_{BH_2}$ be associated with the $\pm$ sign denoting redshift and blueshift. We emphasize that this represents the total frequency shift, such that 1st part gives the gravitational contribution and the 2nd part provides the kinematic component, respectively. Consequently, the interesting region can be defined by radii $r$ exceeding $r_{ISCO}$, where the stable bound orbits exist.
 \begin{figure}
 \includegraphics[width=16pc]{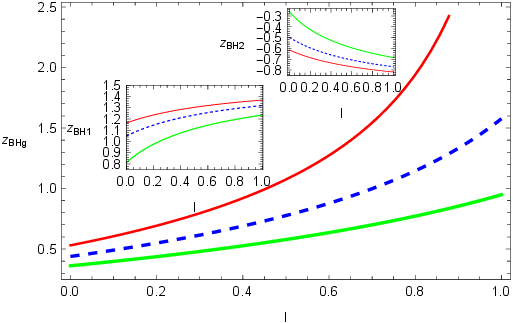}
 \caption{\label{fig3} The redshift curves in bumblebee AdS BH with parameters $m_0=1$ and $\Lambda=0.019,0.039,0.069$.}
\end{figure}
 Thus, by putting Eqs. (\ref{B15C}), (\ref{B15D}) and (\ref{B25}) into Eq. (\ref{B25B}),  we compute the total frequency shift as
\begin{equation}\label{B26c}
1+z_{B H_{1,2}}= \frac{\sqrt{r}\pm \sqrt{\frac{2 r \left(\Lambda  (l+1) r^3-3 {m_0}\right)}{\Lambda (l+1) r^3+6 {m_0}-3 r}}}{\sqrt{3 {m_0}-r}},
\end{equation}
identify
\begin{equation}\label{B28}
1+z_g=\sqrt{\frac{r}{3 {m_0}-r}},
\end{equation}
and
\begin{equation}\label{B29}
z_{k i n_{ \pm}}=\pm \left(\bigg[\frac{r \left(\Lambda  (l+1) r^3-3 {m_0}\right)}{(3 {m_0}-r) \left(\Lambda  (l+1) r^3+6 {m_0}-3 r\right)}\bigg]^{\frac{1}{2}}-1\right).
\end{equation}
It is noted that the Lorentz-violating factor is bounded by x-ray observational measurements of the supermassive BH in 1H0707-495 with the appropriate limits such that  $l\leq 0.6$ for $N=1$~\cite{l21} and $l\leq 0.225$ for $N=2$~\cite{l22}. 
Finally, the loss of energy manifests as a reduction in wave frequency and a corresponding increase in wavelength, a phenomenon generally referred to as redshift. In contrast, when photons gain energy while descending into a gravitational potential well, this leads to a blueshift.

In Fig.~\ref{fig3}, we illustrate the redshift dynamics with fixed parameters. As $\Lambda$ takes higher values, $z_1$ exhibits a noticeable deviation from the redshift profile associated with the BH. It is worth noting that $z_g$ is always positive and exhibits an increasing trend with respect to the free parameters, $z_2$ remains negative and attains lower values, as the $z_{\mathrm{kin}}$ term in Eq.~(\ref{B26c}) dominates. Hence, by defining $\Xi=1+z_1$ and $\Pi=1+z_2$, we get  
\begin{equation}\label{B30}
\mathcal{X} \equiv\Xi \Pi=\frac{3 r}{\Lambda  (l+1) r^3+6 {m_0}-3 r},
\end{equation}
resulting in an expression for the BH mass in terms of the observable frequency shifts $z_1$ and $z_2$, obtained as 
\begin{equation}\label{B31}
m_0= \frac{3 r {\xi_1 \xi_2}-\Lambda l r^3 {\xi_1 \xi_2}-\Lambda  r^3 {\xi_1 \xi_2}+3 r}{6 {\xi_1 \xi_2}},
\end{equation}
the above equation recovers to the Schwarzschild mass in the appropriate limit when $N$/$l$ vanishes.
The maxima of $z_1$ and $z_2$ for stable circular orbits occur at $r_{\text{ISCO}}$, with frequency shifts represented in the table. From Eq. (\ref{B26c}), we obtain
\begin{equation}\label{B32}
\mathcal{Y} \equiv(\Xi+\Pi)^2 =\frac{4 r}{3 {m_0}-r},
\end{equation}
Now, it is straightforward to show that 
{
\begin{equation}\label{B32A}
\frac{\Xi }{\Pi}=\frac{\sqrt{r \left(\Lambda  r^3(l+1)-3 m_0\right)}+\sqrt{r \left(\Lambda  r^3(l+1)+V_1\right)}}{\sqrt{r \left(\Lambda r^3(l+1)+V_1\right)}-\sqrt{r \left(\Lambda r^3(l+1)-3 m_0\right)}}.
\end{equation}}
Here, $V_1=6 m_0-3 r$, we analyze the relations for \(m_0\) and \(l\) to evaluate the BH mass and length scale in the redshift/blueshift framework. From the above expressions, we obtain
\begin{equation}\label{B33}
m_0=\frac{r(\mathcal{Y}+4)}{3\mathcal{Y}}.
\end{equation}
We emphasize that two distinct frequency shifts occur, corresponding to the extremal value  $m_0$, which is associated with light propagation parallel and antiparallel to the emitter's motion around the bumblebee AdS BH. This represents the redshift (receding source) and blueshift (approaching source) signatures, respectively. The maximum and minimum frequency shifts are attained for emitters whose position vector $\mathbf{r}$ relative to the BH's centre is normal to the detector's line of sight; specifically, we show that in the plane where the radial wave vector component $k^r$ vanishes for a distant observer.
For example, in Bardeen BH, using the original metric, $z_{\text{max}}(r, g) < 0.85$, while the effective metric, $z_{\text{max}}(r, g) < 0.71$. On the other hand, for Hayward BH, the corresponding bounds were  $z_{\text{max}}(r, g) < 0.75$ with the original metric and 
$z_{\text{max}}(r, g) < 0.53$ with the effective metric. Similarly, for the ABG BH,  the values were $z_{\text{max}}(r, g) < 0.90$ under the original metric and  $z_{\text{max}}(r, g) < 0.81$ under the effective metric. These results indicated a substantial discrepancy between the outcomes obtained from the original and the effective metrics \cite{RP1}.  

Ultimately, the domain $\mathcal{D}_M^{BH} = \{(r, z, p)\}$, representing the region where a mass parameter satisfies all the conditions for the existence of circular stable orbits of light emitters, and  the difference between $M_{0}(r, z, p)$ and 
$M_{\text{eff}}(r, z, p)$ remained relatively small. In all three illustrative cases, it was found that $M_{\text{eff}}\left(r, p, z_{\text{max}}(r, p)\right) > M_{0}\left(r, p, z_{\text{max}}(r, p)\right)$ \cite{RP1}.

\section{Black hole strong gravitational lensing images}

{In this section, we investigate the shadow cast by the bumblebee AdS BH. To simulate the shadow numerically, we employ the so-called backward ray-tracing method implemented in the \textsc{Arcmancer} code \cite{zz1}. This approach efficiently generates BH images by initializing photon geodesics at the observer’s location and integrating the geodesic equations \eqref{B18} backward in time. Although realistic BH images are typically illuminated by electromagnetic radiation from nearby accretion flows \cite{zz2}, our focus here is solely on the influence of the BH spacetime on photon propagation. Therefore, rather than modeling an accretion disk, we adopt a celestial sphere as the background light source. We partition the celestial sphere into four quadrants, each assigned a distinct color (green, blue, red, and yellow)~\cite{zz3, zz4, zz5, zz6, zz7, zz8}.} A grid of longitude and latitude lines—marked by adjacent gray lines spaced at \(10^\circ\) intervals—is superimposed on this colored background. The resulting distortions of these colored regions and grid lines in the simulated images directly illustrate the effects of strong gravitational lensing induced by the BH. The black region in each image corresponds to the BH shadow: the set of photon trajectories that fall into the event horizon and thus never reach the observer. Additionally, the white regions arise from light rays originating from a reference point located along the line connecting the BH and the observer, providing a clear visualization of the Einstein ring.

We place the observer in the equatorial plane at a radial distance of \(r = 30\,M\) from the BH. The observer’s local Lorentz frame is defined by the orthonormal tetrad \(\{\hat{e}_t, \hat{e}_r, \hat{e}_\theta, \hat{e}_\phi\}\), which can be expressed in terms of the coordinate basis \(\{\partial_t, \partial_r, \partial_\theta, \partial_\phi\}\) as
\begin{equation}
\label{eq:C1}
\hat{e}_{\hat{\mu}} = e^{\nu}{}_{\hat{\mu}} \, \partial_{\nu},
\end{equation}
where the tetrad components \(e^{\nu}{}_{\hat{\mu}}\) satisfy the orthonormality condition
\begin{equation}
\label{eq:C2}
    g_{\mu\nu} \, e^{\mu}{}_{\hat{\alpha}} \, e^{\nu}{}_{\hat{\beta}} = \eta_{\hat{\alpha}\hat{\beta}},
\end{equation}
with \(\eta_{\hat{\alpha}\hat{\beta}}\) being the Minkowski metric. For a static spacetime, it is convenient to choose a diagonal tetrad of the form
\begin{equation}
\label{eq:C3}
e^{\nu}{}_{\hat{\mu}} = 
\begin{pmatrix}
\zeta & 0 & 0 & 0 \\
0 & A^r & 0 & 0 \\
0 & 0 & A^{\theta} & 0 \\
0 & 0 & 0 & A^{\phi}
\end{pmatrix},
\end{equation}
where \(\zeta\), \(A^r\), \(A^{\theta}\), and \(A^{\phi}\) are real coefficients. Imposing the orthonormality condition yields
\begin{equation}
\label{eq:C4}
\zeta = \frac{1}{\sqrt{-g_{tt}}}, \quad
A^r = \frac{1}{\sqrt{g_{rr}}}, \quad
A^{\theta} = \frac{1}{\sqrt{g_{\theta\theta}}}, \quad
A^{\phi} = \frac{1}{\sqrt{g_{\phi\phi}}}.
\end{equation}
The locally measured photon four-momentum \(k^{\hat{\mu}}\) is determined by projecting the coordinate four-momentum \(k^{\mu}\) onto the observer’s tetrad basis, given by
\begin{equation}
\label{eq:C5}
k^{\hat{t}} = -k_{\hat{t}} = -e^{\nu}_{\hat{t}} k_{\nu}, \qquad
k^{\hat{i}} = k_{\hat{i}} = e^{\nu}_{\hat{i}} k_{\nu} \quad (i=r,\theta,\phi).
\end{equation}
In the spacetime of the bumblebee AdS BH \eqref{B77}, this projection yields
\begin{eqnarray}
\label{eq:C6}
k^{\hat{t}} &=& \frac{1}{\sqrt{-g_{tt}}} \, E_\gamma, \qquad \qquad
k^{\hat{r}} = \frac{1}{\sqrt{g_{rr}}} \, k_{r}, \nonumber\\
k^{\hat{\theta}} &=& \frac{1}{\sqrt{g_{\theta\theta}}} \, k_{\theta}, \qquad \qquad
k^{\hat{\phi}} = \frac{1}{\sqrt{g_{\phi\phi}}} \, L_\gamma,
\end{eqnarray}
where \(E_\gamma\) and \(L_\gamma\) are the conserved energy and axial angular momentum of the photon, respectively. We determine the apparent position \((x, y)\) of the photon in the observer’s sky, the following hold \cite{zz3, zz4, zz5, zz6, zz7, zz8}
\begin{eqnarray}
\label{eq:C7}
x &=& -r_{\text{obs}} \frac{k^{\hat{\phi}}}{k^{\hat{r}}}
= -r_{\text{obs}} \frac{L_z}{\sqrt{g_{rr} g_{\phi\phi}} \dot{r}}, \nonumber\\
y &=& r_{\text{obs}} \frac{k^{\hat{\theta}}}{k^{\hat{r}}}
= r_{\text{obs}} \frac{\sqrt{g_{\theta\theta}} \dot{\theta}}{\sqrt{g_{rr}} \dot{r}}.
\end{eqnarray}
The simulations are performed at a resolution of \(800 \times 800\) pixels. For each pixel, we integrate the corresponding null geodesic backward in time until the photon either reaches the celestial sphere or crosses the event horizon. 
\begin{figure}[ht]
    \centering
    \includegraphics[width=8cm]{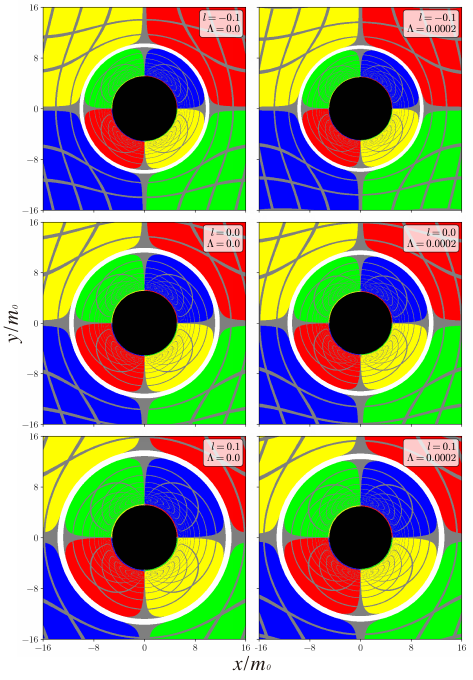}
    \caption{Strong gravitational lensing image of a bumblebee AdS BH as seen on the celestial sphere. Here we set the mass parameter $m_0 = 1$, $r_{obs} = 30M$ and $\theta_{obs} = \pi/2.$}
    \label{fig:shadow}
\end{figure}
Fig.~\ref{fig:shadow} presents the resulting BH shadow images for various combinations of the bumblebee parameter \(l = -0.1, 0, 0.1\) and the cosmological constant \(\Lambda = 0, 0.0002\). The shadow itself appears circular in all cases, with only minor deviations induced by the parameters \(l\) and \(\Lambda\). This insensitivity reflects the fact that the photon capture region is only weakly dependent on these parameters. However, the influence of \(l\) on the strong gravitational lensing features outside the shadow is striking. As \(l\) increases, the bright white Einstein ring expands noticeably, and the gray grid lines of the celestial sphere are stretched outward, indicating enhanced light bending at larger angular separations from the axis. In contrast, a small positive cosmological constant \(\Lambda = 0.0002\) produces the opposite effect, both the Einstein ring and the lensed grid lines contract slightly compared to the \(\Lambda = 0\) case.  

These simulations suggest that although \(l\) and \(\Lambda\) have a negligible impact on the shadow size, their effects become increasingly pronounced on larger angular scales. This behavior is consistent with the expectation that even small deviations in \(l\) and \(\Lambda\) can induce modifications to the metric that grow as \(\mathcal{O}(r^2)\) \eqref{B77}, leading to enhanced light deflection at large observer distances. 
Thus, high-resolution imaging of lensed background structures may present a more sensitive analyze of bumblebee and cosmological constant effects than measurements of the shadow size alone.

\section{Summary and conclusions}

In this study,  we have extended a general relativistic framework \cite{l11,l12,l13}  to the bumblebee AdS BH in order to investigate the geodesic structure of massive particle motion, and the corresponding ray-tracing signatures. Furthermore, we have investigated a particular utilization of bumblebee AdS BH that is connected to the Schwarzschild solution, providing an observable example within this general relativistic framework.
We examined the dependence of the frequency shifts on both peculiar and gravitational redshifts, showing that the total redshift increases with the receding peculiar velocity of the BH, whereas the corresponding blueshift decreases.

Notably, there exists a zero point in the redshift/blueshift at a specific velocity, where the kinematic contribution becomes significant. The stable region is determined by $r>r_h\ \text{and} \ V_r''(r)>0$. However, in the Schwarzschild limit $\Lambda=0$, the expression (\ref{Bisco}) reproduces  to the well-known result $r_{\rm ISCO}=6m_0$.
We compare our results with those reported in recent publications \cite{C1,C2,C3}, such as the derived metric constituents obtained for Bardeen, Hayward, and ABG BHs. We highlight that this general relativistic framework has been formulated for numerous Schwarzschild BH that are situated within the central regions of various active galactic nuclei, such as NGC 4258 \cite{RD16} and TXS-2226-184 \cite{C5}, respectively. 

In addition, we made a quantitative analysis of the deformation of the BH shadow induced by the parameters \(l\) and \(\Lambda\) effects. Some interesting quantitative results are presented in Fig.~\ref{fig:shadow}. In our subsequent study, we have computed the mass of the polymerized BH residing in different active galactic nuclei and assessed the quantum corrections using the general relativistic framework.
It is anticipated that future studies will further explore the upper boundary behavior of \(r_{\mathrm{ISCO}}\) under various  matter-field configurations to obtain a more comprehensive understanding of ISCO dynamics in modified gravity frameworks.

In summary, our findings impose a fundamental limit on the location of stable particle orbits around BHs, thereby connecting theoretical predictions with observable phenomena. This study emphasizes the close relationship between BH geometry offering a foundation for more comprehensive analyses of compact-object dynamics in both standard and extended theories of gravity.

\section*{Acknowledgements}
The work was sponsored by the National Natural Science Foundation of China (Grant Numbers: 12235007, 12375003). The work is also supported by NSFC Grant No. 12547132. 

\section*{Conflict of Interest} 
The authors declare that they have no conflicts of interest.
 \section*{Data Availability Statement} Data will be made available on request.

\end{document}